\documentclass[aps,twocolumn,superscriptaddress,english,10pt,nofootinbib]{revtex4}

\usepackage{float}

\usepackage{amssymb, amsmath, bm, dcolumn, epsf, graphicx, latexsym, slashed, simplewick}
\usepackage[utf8]{inputenc}
\usepackage[normalem]{ulem}
\usepackage{amssymb}

\usepackage{float}

\usepackage{color}

\def\be{\begin{equation}}
\def\ee{\end{equation}}
\def\bea{\begin{eqnarray}}
\def\eea{\end{eqnarray}}

\usepackage{hyperref}
\usepackage{comment}
\usepackage{soul}

\usepackage[normalem]{ulem}

\begin{document}


\title{Quantum Gravity Signatures in the Late-Universe}

\author{Michael W.~Toomey}
\affiliation{ Department of Physics, Brown University, Providence, RI 02912-1843, USA}
\affiliation{ Brown Theoretical Physics Center, Brown University, Providence, RI 02912-1843, USA}

\author{Savvas Koushiappas}
\affiliation{ Department of Physics, Brown University, Providence, RI 02912-1843, USA}
\affiliation{ Brown Theoretical Physics Center, Brown University, Providence, RI 02912-1843, USA}

\author{Bruno Alexandre}
\affiliation{Theoretical Physics Group, The Blackett Laboratory, Imperial College,\\
Prince Consort Rd., London, SW7 2BZ, UK}

\author{Jo\~{a}o Magueijo}
\affiliation{Theoretical Physics Group, The Blackett Laboratory, Imperial College,\\
Prince Consort Rd., London, SW7 2BZ, UK}

\begin{abstract}
    We calculate deviations in cosmological observables as a function of parameters in a class of connection--based models of quantum gravity. In this theory  non-trivial modifications to the background cosmology can occur due to a distortion of the wave function of the Universe at the transition from matter to dark energy domination (which acts as a ``reflection'' in connection space).
    We are able to exclude some regions of parameter space and show with projected constraints that future experiments like DESI will be able to further constrain these models. An interesting feature of this theory is that there exists a region of parameter space that could naturally alleviate the $S_8$ tension. 
\end{abstract}

\maketitle

\section{Introduction}

One of the most striking features of $\Lambda$CDM cosmology is the recent transition from decelerated to accelerated expansion. 
While the dynamics of this transition through the lens of the concordance model is well understood and has been studied \textit{ad nauseam}, it has recently been pointed out \cite{Alexandre:2022ijm} that from the perspective of some theories of quantum gravity, such a transition can result in a novel evolution of the background cosmology. The reason is because the expansion rate $\dot{a}$ directly maps to the (connection or gauge field) degree of freedom of the gravitational theory and the semi-classical limit at the transition breaks down. This manifests as a brief enhancement in the Hubble parameter around dark energy — matter equality. 

This feature is restricted to first order gravity, a connection based formulation of general relativity where the metric is no longer a fundamental degree of freedom. This “connection-based” representation of general relativity  is classically entirely equivalent to the metric formulation but can have several advantages over second order gravity, e.g., coupling gravity to fermions, a feature that played a pivotal role in the development of loop quantum gravity. 
Here, we are interested in the effects of the recent dark energy—matter transition in the context of the connection representation  because of the predictive effects on the dynamical evolution of the connection variable, $b = \dot{a}$.
Classically, $b$ decreases before the transition (decelerated expansion, $\dot b=\ddot a<0$), reaches a minimum, and then increases (accelerated expansion, $\dot b=\ddot a>0$). 
During the dark energy—matter transition, interference of incident and reflected waves manifests a brief enhancement in the Hubble parameter, but more importantly, still, the wave function is squashed against the minimum-$b$-wall, creating a skewed wave function, and so a bias towards larger Hubble parameters \cite{Alexandre:2022ijm}.

While its possible to directly constrain deviations to the background cosmology, if modifications of this nature are large enough they can also appreciably alter the growth of large-scale structure (LSS) and low-$\ell$ features in the cosmic microwave background (CMB) via the integrated Sachs-Wolfe (ISW) effect. In this work, we explicitly calculate the novel signatures present in this class of quantum gravity models.

In Section \ref{Quant} we review the model of quantum cosmology considered in this work, in Section \ref{CosTh} the relevant cosmological signatures and in Section \ref{DnC} we discuss the implication of our results. 

\section{Theory of Quantum Cosmology}\label{Quant}

The model studied here is a generalization of unimodular gravity \cite{PhysRevD.40.1048,Smolin:2009ti,PhysRevD.43.3332,Daughton1993InitialCA,sorkin1,sorkin2} based on \cite{HENNEAUX1989195}. In this formulation the cosmological constant is no longer a coupling constant, but rather an integration constant, or a constant as a result of the equations of motion: a constant on-shell only. Since $\Lambda$ appears in the phase space (rather than being a parameter) it has a dual variable, which acts as a physical time variable (this turns out to be unimodular or 4-volume time on-shell~\cite{Bombelli,UnimodLee2})
 As a result, the wave function of the Universe in these theories does not obey the standard Wheeler-DeWitt equation, but rather a Schr\"{o}dinger-like equation.  One benefit of this approach is that one is not faced with the same normalization issues that typically plague solutions to the Wheeler-DeWitt equation \cite{Witten:2003mb}\footnote{Recent progress suggest this problem may actually be surmountable \cite{Alexander:2022ocp}.} since we can now superpose waves with different $\Lambda$ to form wave packets.
 
 Critically, this approach also avoids the issue of time in quantum gravity \cite{Isham:1992ms,Kuchar:1991qf}, because as already mentioned, the canonical conjugates to the integration constants in these theories are great candidates for a cosmic time. For example, the cosmological constant is conjugate to the 4D volume of past observers. This approach can also be extended more generally to constants derived from a perfect fluid which will also have a conjugate corresponding to a cosmic time. When one fluid dominates over the other, we can think of that fluid's conjugate serving the role of a clock for the Universe. We will now outline some of the details of the model considered in this work. For more details see \cite{Alexandre:2022ijm}.

\subsection{Classical Theory}
The Einstein-Cartan action in the connection based formulation and a homogeneous and isotropic minisuperspace model is given by,
\begin{equation}
    S_{GR} = \frac{3V_c}{8\pi G} \int dt \left( \dot{b}a^2 + N a (b^2 + k) \right),
\end{equation}
where the connection variable  is $b = \dot{a}/N$, $N$ is the lapse function, $V_c = \int d^3x$ is the coordinate 3-volume from the slicing and $k = 0, \pm 1$ is the standard curvature term. In this paper we are interested in the effects at the transition from matter to dark energy domination. Thus, we consider a cosmological model for two perfect fluids with equations of state $w = 0$ and $w = -1$. A perfect fluid in the minisuperspace representation is defined at the level of the action as \cite{Brown:1992kc,PhysRevD.41.1125},
\begin{equation}
    S_{\rm fl} = \int dt \left( U\dot{\tau} - N a^3 V_c \rho n \right)
    \label{fld_act}
\end{equation}
with $U$ the total (conserved) particle number,  $n = U / a^3 V_c$ is the number density, and $\tau$ a Lagrange multiplier. Assuming the standard evolution of a fluid with equation of state $w$ one can define a new conserved variable
\begin{equation}
    m = \frac{8\pi G \rho_0}{3 V_c}\left( \frac{U}{V_c} \right)^{1 + w}
\end{equation}
from which we can write down an alternative form of Eq.~\ref{fld_act} (see also \cite{Gielen:2016fdb,Gielen:2020abd,Gielen:2021igw})
\begin{equation}
    S^{(w)}_{\rm fl} = \frac{3 V_c}{8\pi G} \int dt \left( \dot{m}\chi - N \frac{m}{a^{3w}} \right).
\end{equation}
With this in hand we can now write down our full action for gravity with dark matter and dark energy,
\begin{equation}
\begin{split}
        S_{GR} = \frac{3V_c}{8\pi G} \int dt ~ \bigg[ ~ & \dot{b}a^2 
        + \dot{m}\chi_1 + \dot{\Lambda}\chi_2 \\ 
        - N a & \left( - (b^2 + k) + \frac{m}{a} + \frac{\Lambda}{3}a^2 \right) \bigg].
\end{split}
\end{equation}
This action implies the corresponding set of non-vanishing Poisson brackets,
\begin{equation}
    \{b,a^2\} = \{m, \chi_1\} = \{\Lambda, \chi_2\} = \frac{8\pi G}{3V_c},
    \label{PoissB}
\end{equation}
where we can view $b$ as a coordinate and $a^2$ as it conjugate momentum. Additionally, we have our ``constants'' $m$ and $\Lambda$ along with their conjugate time variable. Now we can write down the Hamiltonian density for the theory,
\begin{equation}
    \mathcal{H} = \frac{3V_c}{8\pi G} N a \left( - (b^2 + k) + \frac{m}{a} + \frac{\Lambda}{3}a^2 \right).
\end{equation}
Hamilton's equations for $N$ give the Hamiltonian constraint,
\begin{equation}
    -\left(b^2 + k \right) + \frac{m}{a} + \frac{\Lambda}{3}a^2 = 0,
    \label{fmann}
\end{equation}
which is just the Friedmann equation. Going forward we will now make two simplifications. First we will abandon the variables for dark energy in favor of a canonically transformed pair, $\phi = 3/\Lambda$ and T$_\phi = -3 \chi_2/\phi^2$, and second we will define a potential $V(b) = (b^2 + k)$.
As demonstrated in \cite{MAGUEIJO2021136487,Magueijo:2021pvq,Gielen:2022dhg,Alexandre:2022ijm} the Hamiltonian constraint can then be recast in the form of two constraints,
\begin{equation}
    h_\pm(b)a^2 - m^2  = 0,
    \label{constraint}
\end{equation}
where one will hold before the bounce and the other afterwards.

\begin{figure*}[!t]
    \centering
    \includegraphics[width=\linewidth]{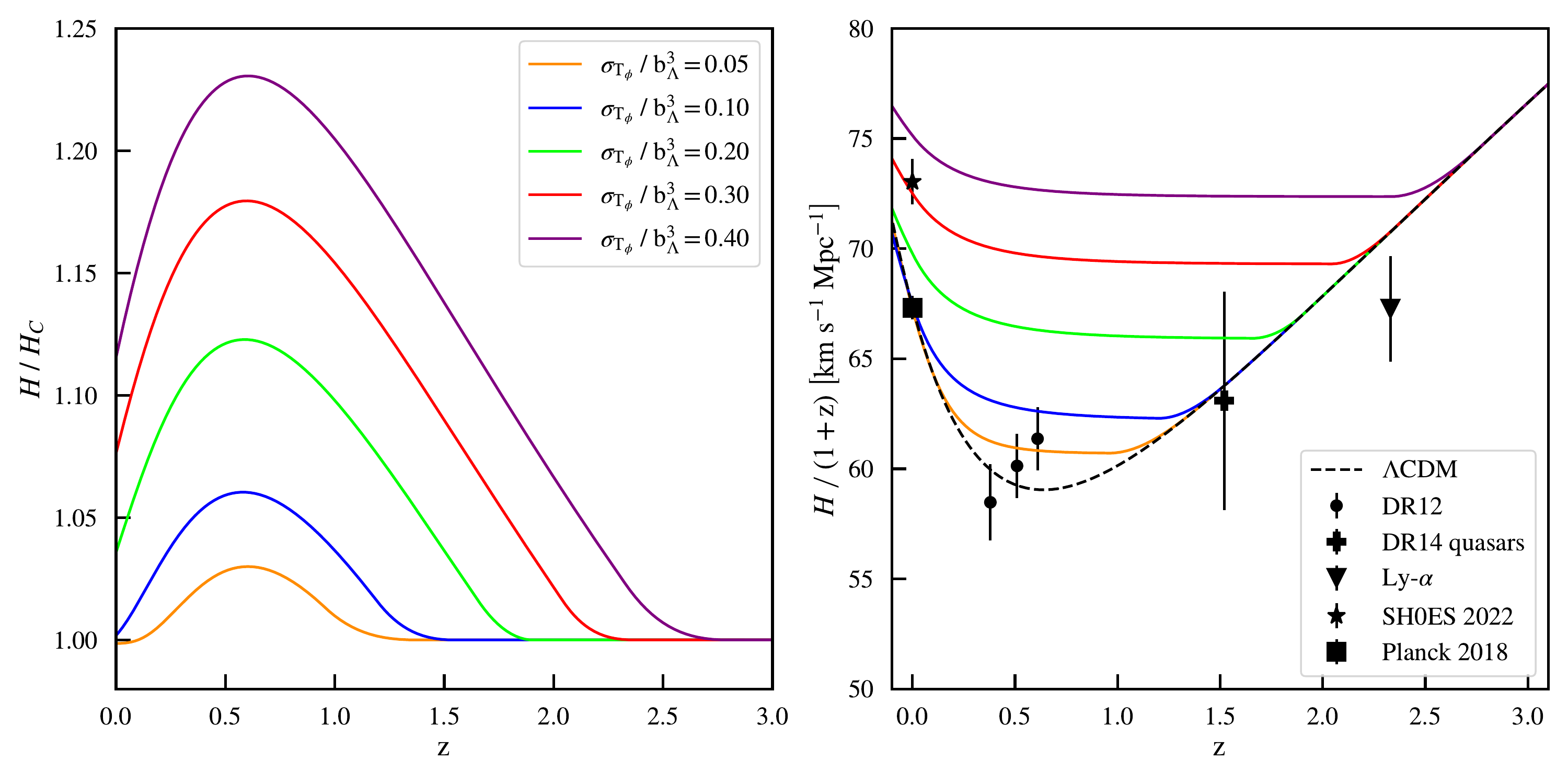}
    \caption{\textit{Left:} The fractional change in the Hubble parameter from the base $\Lambda$CDM model, with best-fit Planck 2018 parameters from Eq.~\ref{LCDMparam}, for different parameter values of the quantum cosmology model. \textit{Right:} Evolution of the expansion rate for cosmologies considered in this work along with various experimental constraints.}
    \label{fig:Hub}
\end{figure*}

\subsection{Quantum Theory}

We can now quantize the theory by promoting the classical Poisson bracket (Eq.~\ref{PoissB}) to a commutation relation,
\begin{equation}
    \left[\hat{b}, \hat{a}^2 \right] = i \frac{l^2_P}{3V_c}
\end{equation}
with the reduced Planck length $l_p = \sqrt{8\pi G_N \hbar}$. Making the choice of representation that diagnolizes $\hat{b}$ we can write its conjugate operator as,
\begin{equation}
    \hat{a}^2 = - i \mathfrak{h} \frac{\partial}{\partial b}
    \label{opa2}
\end{equation}
with $\mathfrak{h} = l^2_P / 3 V_c$ the ``effective Planck parameter'' \cite{Barrow:2020coo}. We now have ingredients to define a two-branch first order formulation of a quantum theory of gravity after combining Eqs.~\ref{constraint} and \ref{opa2},
\begin{equation}
    \left[-i \mathfrak{h} h_\pm(b)  \frac{\partial}{\partial b}- m^2 \right] \psi = 0,
\end{equation}
which is a Wheeler-DeWitt equation for the connection represtation. To generalize the Chern-Simons functional the following linearizing variable is defined,
\begin{equation}
    X_\pm(b) = \int \frac{d b}{h_\pm (b)}.
\end{equation}
The general solution is a superposition of different valued constants $\boldsymbol\alpha = (m^2,\phi)$ for spatial monochromatic functions $\psi_s(b;\boldsymbol\alpha)$ which are normalized such that $|\psi_s|^2 = 1/(2\pi\mathfrak{h})^D$ and the total solution is of the form,
\begin{equation}
    \psi(b,\mathbf{T}) = \int d\boldsymbol{\alpha}\mathcal{A}(\boldsymbol{\alpha}){\rm exp}\left[ - \frac{i}{\mathfrak{h}} \boldsymbol{\alpha}\cdot\mathbf{T} \right] \psi_s(b;\boldsymbol{\alpha}),
\end{equation}
where $\mathbf{T}$ is the conjugate time to $\boldsymbol\alpha$ and $D$ is the number of conserved parameters $\boldsymbol\alpha$. Here $\psi_s$ is made up of an incident and a reflected wave, associated with $X_\pm$.

Since we would want to reproduce a semiclassical limit at late times, the choice for $\mathcal{A} = \sqrt{\mathcal{N}(\alpha_{i0},\sigma_{\alpha_i})}$ corresponding to wave packets with Gaussian aplitudes
\begin{equation}
    \mathcal{A}(\alpha_i) = \frac{1}{\left(2\pi \sigma_i^2\right)^{1/4}} {\rm exp} \left[ - \frac{(\alpha_i - \alpha_{i0})^2}{4\sigma_i^2} \right]
    \label{packet}
\end{equation}
leads to wave packets of the form,
\begin{equation}
    \psi_{\pm i}(b,T_i) = \frac{1}{(2\pi \sigma^2_{T_i})^{1/4}} {\rm exp} \left[ - \frac{\left( X^{\rm eff}_{\pm i}(b) - T_i \right)^2}{4\sigma_{T_i}^2} \right]
\end{equation}
where $\sigma_{{\rm T}_i} = \mathfrak{h}/2\sigma_i$ and $X^{\rm eff}_{\pm i}(b)$ is an effective functional adapted to wave packets in the saddle point approximation (see~\cite{Magueijo:2021pvq,Gielen:2022dhg,Alexandre:2022ijm} for details). The probability for a given $b$ at $T_i$ in the semi-classical limit can be found to be
\begin{equation}
    \mathcal{P}(b,T_i) = \left| \frac{d X^{\rm eff}_{+ i}}{d b} \right| \left| \psi_+ \right|^2 + \left| \frac{d X^{\rm eff}_{- i}}{d b} \right| \left| \psi_- \right|^2 
\end{equation}
(see~\cite{Gielen:2022dhg,Alexandre:2022ijm} for more details; no interference terms appear in the semiclassical approximation). 
From this, one can then calculate modifications induced in the Hubble parameter as,
\begin{equation}
    \Bar{b}(T_\phi) = \int_{b_\Lambda}^\infty db ~ \mathcal{P}(b,{\rm T}_\phi) b, 
\end{equation}
where $b_\Lambda$ is the value at dark energy--matter equality and the contribution to the Hubble parameter originating from quantum effects is explicitly given as,
\begin{equation}
    H_{\rm Q} =  \frac{\Bar{b} - b}{a}
\end{equation}
such that we can think of the Hubble parameter as the sum of the classical background solution and this new quantum addition: $H = H_{\rm C} + H_{\rm Q}$.

\begin{figure*}[!t]
    \centering
    \includegraphics[width=\linewidth]{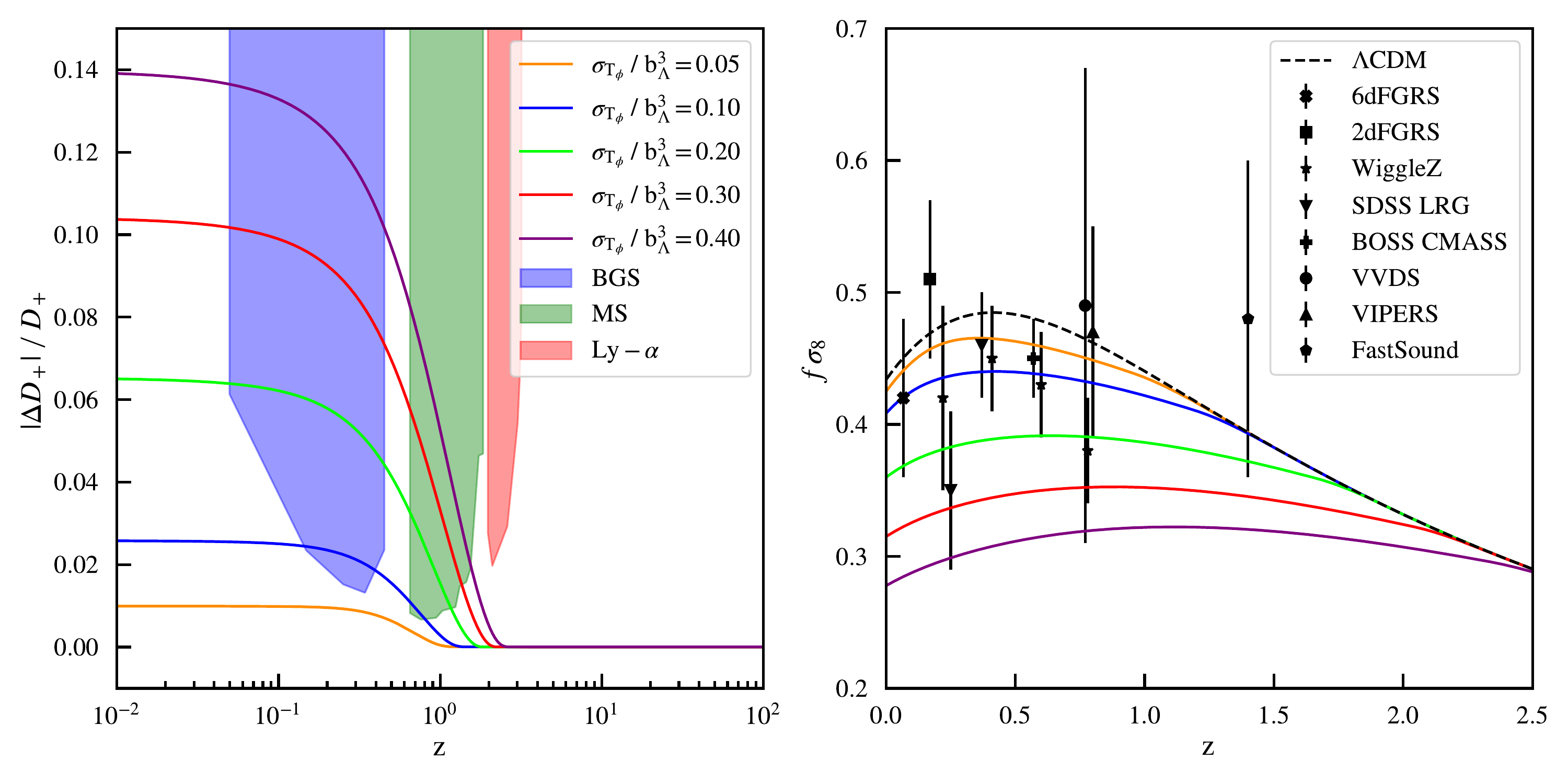}
    \caption{\textit{Left:} Deviation in the growth factor from the base $\Lambda$CDM cosmology with projected contraints from DESI. \textit{Right:} Evolution of $f\sigma_8$ for $\Lambda$CDM (dashed), with parameters from Eq.~\ref{LCDMparam}, and for the quantum cosmology model with differing values of $\sigma_{{\rm T}_\phi}$ (solid). Over plotted are constraints on $f\sigma_8(z)$ from RSD measurements from galaxy surveys.}
    \label{fig:growth_d}
\end{figure*}

\section{Cosmological Imprints}\label{CosTh}

Deviations from the standard $\Lambda$CDM expansion history in the late-Universe will impact the growth of structure and can be constrained from observational probes ranging from measurements of the growth factor, $f \sigma_8$, and the late-ISW effect. However, before we can calculate quantum effects our first step is to calculate the standard, classical $\Lambda$CDM cosmology. For this we use the best-fit Planck 2018 parameters given by \cite{Planck:2018nkj},
\begin{align}
     \label{LCDMparam}
     H_0&=67.32 \, {\rm km/s/Mpc} &   100\omega_b&=2.283, \\
     \omega_{\rm cdm}&=0.1201 &10^9 A_s &= 2.210, \nonumber \\  n_s&=0.9661 &    \tau_{\rm reio}&=0.0543, \nonumber
\end{align}
to calculate the cosmology with the Boltzmann code \texttt{CLASS} \cite{2011JCAP...07..034B}. Separately, we then calculate the expected deviations in the expansion rate due to quantum effects in the very late-Universe for a range of values for $\sigma_{{\rm T}_\phi}$. Since the $\Lambda$ clock becomes unreliable as we enter deeper into matter domination, we smoothly interpolate between the quantum effects at the bounce and the classical solution at higher redshifts, i.e. $H_{\rm Q} \rightarrow 0$.  This is to be expected since the relevant semi-classical wave function will correspond to the matter clock as we move to higher redshifts and away from the bounce. 

In the left panel of Fig.~\ref{fig:Hub} we show the evolution of the fractional change in the Hubble parameter for a set of different values of $\sigma_{{\rm T}_\phi}/b_\Lambda^3$, while in the right panel of Fig.~\ref{fig:Hub} we show the evolution of  $H / (1 + z) = \dot{a}$, i.e. the expansion rate, against direct BAO measurements from BOSS DR12 \cite{BOSS:2016wmc}, BOSS DR14 quasars \cite{Zarrouk:2018vwy}, and BOSS Ly-$\alpha$ \cite{Bautista:2017zgn} as well as the best fit $H_0$ values from SH0ES \cite{Riess:2021jrx} and for Planck 2018 \cite{Planck:2018nkj}.  Notice that the deviation from the classical evolution in the Hubble parameter becomes more pronounced for larger values of $\sigma_{{\rm T}_\phi}/b_\Lambda^3$, which also results in a characteristic flattening of the evolution of the expansion rate. This effect can be understood as directly stemming from the increase in variance for the wave packets which results in a more pronounced skewness of the wave function. In turn, this lends itself to a bias for larger values of the Hubble parameter and an impact on the cosmology for a larger range of redshifts around the bounce. 

It is worth noting that the region of parameter space suggested in \cite{Alexandre:2022ijm} that could alleviate the Hubble tension ($\sigma_{{\rm T}_\phi}/b_\Lambda^3 \approx 0.30$) is not favored (due to the BOSS DR12 measurements).

\subsection{Signatures in the matter power spectrum}

Beyond modifications at the background level, there will also be an impact on the evolution of perturbations. Changes in the abundance and evolution of the various constituents of the Universe is readily encoded in the matter power spectrum. Current and future probes of the matter power spectrum can then be leveraged to distinguish between different models of dark matter, extensions of general relativity, and dark energy.

The growth of linear perturbations can be derived from the time evolution of the growth factor, 
\begin{equation}
    D''(\tau) = - a H D'(\tau) + \frac{3}{2} a^2 \rho_M D(\tau), 
    \label{groth_eom}
\end{equation}
where primes denote derivaties with respect to conformal time. 
To establish the sensitivity of the cosmology from varying $\sigma_{{\rm T}_\phi}$ we have plotted the deviation in the growth factor from $\Lambda$CDM in Fig.~\ref{fig:growth_d}. We also include the projected constraints for the Dark Energy Spectroscopic Instrument (DESI) \cite{DESI:2016fyo} for changes in the growth factor relative to $\Lambda$CDM from the Bright Galaxy Survey (BGS) at very low redshifts, with the Main Survey (MS) around z $\approx$ 1, and at higher redshifts with the Lyman-$\alpha$ survey (Ly-$\alpha$). As we anticipate, before dark energy--matter equality the evolution is identical to the classical expectation until the bounce where the semi-classical limit breaks down and we see values of $\sigma_{{\rm T}_\phi}/b_\Lambda^3 \gtrsim 0.10$ are likely to be constrained by DESI.

Another important probe of new physics is the growth rate of perturbations, $f \equiv d \log{D} / d \log{a}$, which can be measured directly with supernova (but only for very low redshifts) and with galaxies surveys like BOSS through redshift-space distortions (RSD) \cite{10.1093/mnras/227.1.1}. For the latter, this stems from distortions along the line-of-sight in the distribution of galaxies (in redshift space) stemming from an induced peculiar velocity as they are drawn toward the center of potential wells. This can be measured from the galaxy power spectrum which can be decomposed as, 
\begin{equation}
    P_{\rm g}(k,\mu_k,z) = P(k,z) \left(b_1 + f \mu_k^2\right)^2 + P_N, 
    \label{galpow}
\end{equation}
where  $P(k,z)$ is the standard linear matter power spectrum, $b_1$ is the galaxy bias, $\mu_k = \hat{\mathbf{e}}_z \cdot \hat{\mathbf{k}}$ the angle between the wavevector and line-of-sight, and $P_N$ is shot noise. From Eq.~\ref{galpow} it is clear that its not possible to disentangle the growth rate from the matter power spectrum, thus the actual observable from galaxy surveys is the redshift dependant quantity $f\sigma_8$ where $\sigma_8$ can be thought of as the amplitude of the power spectrum but is defined as the RMS linear-theory \textit{mass} fluctuation in a sphere of radius 8~$h^{-1}$Mpc. In k-space this is calculated as an integral over the matter power spectrum,
\begin{equation}
    \sigma_8(z)^2 = \int d \log{k} ~ W(kR)^2 k^3 P(k,z),
\end{equation}
where $W(kR)$ is a window function of scale 8~$h^{-1}$Mpc.

In Fig.~\ref{fig:growth_d} we plot the evolution of $f\sigma_8(z)$ relative to constraints from RSD measurements from  BOSS DR11 \cite{2013MNRAS.429.1514S}, WiggleZ \cite{2011MNRAS.415.2876B}, SDSS LRG \cite{2012MNRAS.420.2102S}, VVDS \cite{Guzzo:2008ac}, VIPERS \cite{delaTorre:2013rpa}, FastSound \cite{Okumura:2015lvp}, 2dFGRS \cite{2dFGRS:2004cmo}, and 6dFGRS \cite{2012MNRAS.423.3430B}. As we would expect from a model that results in an enhancement in the Hubble parameter, there is a corresponding suppression in the growth of structure relative to the baseline $\Lambda$CDM model. This is rather unique among models that make modifications in the gravitational sector as typically this leads to an enhancement in the growth rate, e.g. scalar-tensor theories.
For the curves in Fig.~\ref{fig:growth_d} we calculate the $\chi^2$ values for the plotted data points. For $\Lambda$CDM we find that $\chi^2 \sim 14$ and we find for $\sigma_{{\rm T}_\phi}/b_\Lambda^3 = 0.05,~0.10,~0.20$ that $\chi^2 \sim 9,~7,~19$ respectively. This implies that there could be a region in parameter space that improves the fit between data and model for $0 \lesssim \sigma_{{\rm T}_\phi}/b_\Lambda^3 \lesssim 0.20$ but more careful data analysis would be required to make a more definitive statement. Quantifying in terms of the $S_8$ parameter ($S_8 \equiv \sigma_8 \sqrt{\Omega_m /0.3}$) for $\sigma_{{\rm T}_\phi}/b^3_\Lambda = 0.05, 0.10, 0.20$, this results in a suppression $\sim 3$\%, 5\%, and  12\% respectively. Interestingly, the suppression in this region of parameter space is enough to help alleviate the $S_8$ tension \cite{Abdalla:2022yfr}. 

\subsection{Signatures in the CMB}

\begin{figure}[!t]  
\includegraphics[width=\linewidth]{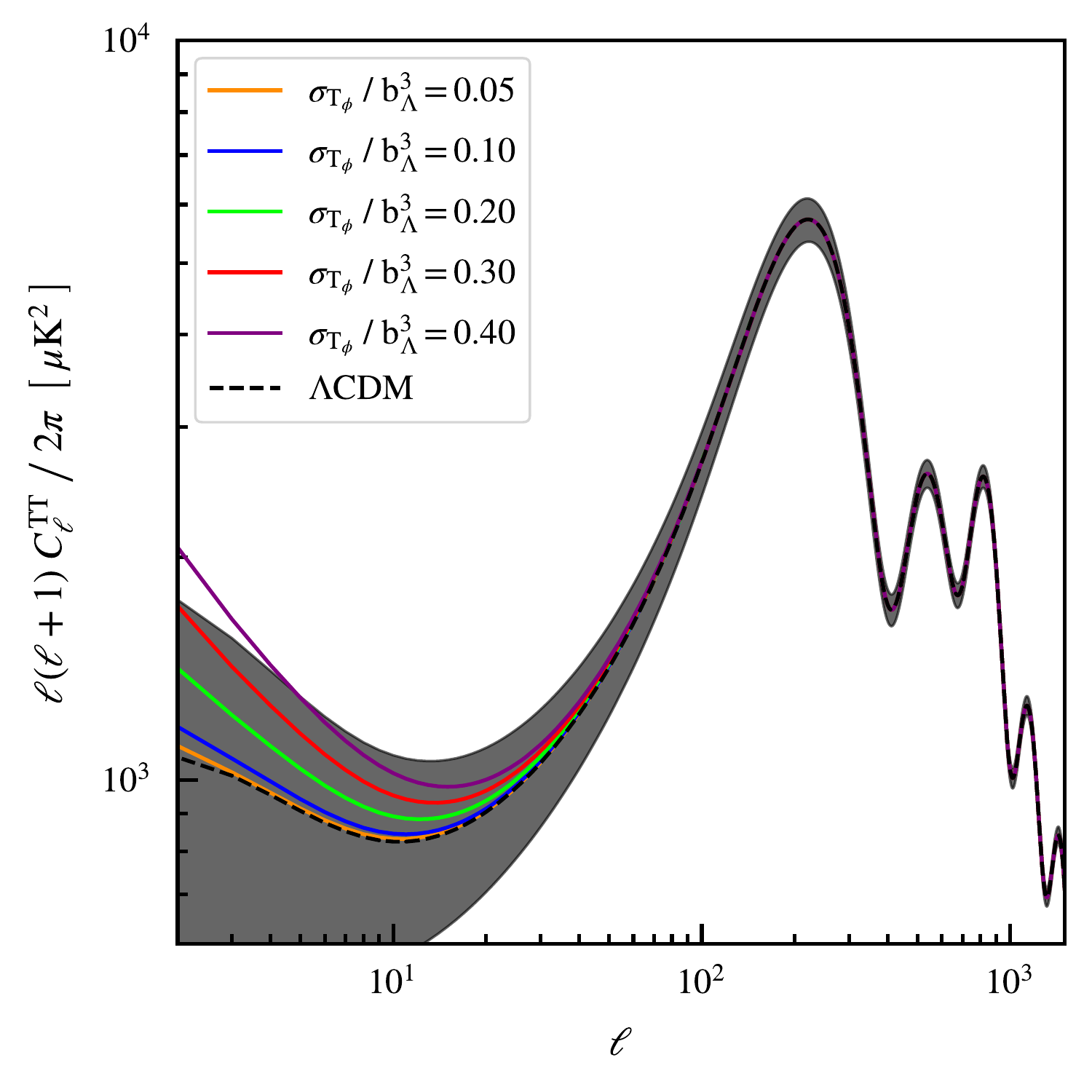}
    \caption{ CMB temperature power spectrum including contribution from the late-ISW effect for $\Lambda$CDM (dashed), with parameters from Eq.~\ref{LCDMparam}, and for the quantum cosmology model with differing values of $\sigma_{{\rm T}_\phi}$ (solid). The gray band corresponds to cosmic variance.}
    \label{fig:ISW_n_fS8}
\end{figure}

While CMB power spectra encode rich information about the physics around recombination, it is also a probe of other processes at intervening epochs leading up to today. Relevant to the model considered here, changes in the growth of structure in the late-Universe can be probed via the late-ISW effect \cite{Hu:2001bc} defined as,
\begin{equation}
    \Theta^{\rm ISW}(\hat{\mathbf{n}}) = 2 \int^{\tau_0}_{\tau_{\rm dec}} d\tau ~\Phi',
\end{equation}
which is an integral from recombination to today over the rate of change of the potential and can serve as a sensitive probe of new physics beyond the $\Lambda$CDM model. The angular power spectrum for the late-ISW effect is given by,
\begin{equation}
    C^{\rm ISW}_\ell = \frac{8}{\pi} \int d \log k~k^3  \left< \left| \int d\tau ~\frac{\partial \Phi_\textbf{k}}{\partial \tau} j_\ell (k\tau) \right|^2\right>,
    \label{ISWpower}
\end{equation}
where we have transitioned to Fourier space and $j_\ell(k\tau)$ is the spherical Bessel function. The Poisson equation in k-space gives the Newtonian potential which, ignoring the velocity perturbations, can be written as
\begin{equation}
    \Phi_\textbf{k} = \frac{3 H_0^2 \Omega_m}{2 k^2} \frac{D}{a} \delta(k).
    \label{Phi}
\end{equation}
Eq.~\ref{Phi} can then be differentiated with respect to conformal time to get,
\begin{equation}
     \frac{\partial \Phi_\textbf{k}}{\partial \tau} = \frac{3 H_0^2 \Omega_m}{2 k^2} \delta(k) \left( \frac{\partial D}{\partial a} \dot{a} - \frac{D}{a} H \right),
     \label{diffPhi}
\end{equation}
and then be combined with Eqs.~\ref{ISWpower} and \ref{diffPhi} to yield the expression of the full angular power spectrum,
\begin{equation}
C^{\rm ISW}_\ell = \frac{18}{\pi} H_0^4 \Omega_m^2 \int \frac{d \log k}{k} P(k)  \left[ \int d\tau  \mathcal{H} D\left( f - 1 \right) j_\ell (k\tau) \right]^2,
\label{lISW}
\end{equation}
where $P(k)$ is the matter power spectrum. Having solved for the background cosmology and evolution of perturbations it is then straightforward to numerically calculate the late-ISW contribution to the CMB using Eq.~\ref{lISW}. 

In Fig.~\ref{fig:ISW_n_fS8} we show the full temperature power spectrum for different values of $\sigma_{{\rm T}_\phi}$ together with the base $\Lambda$CDM model. 
We see the impact of the quantum effects as an excess of power at low~$\ell$ relative to $\Lambda$CDM. This behavior is easy to understand given the explicit dependence on the Hubble parameter in Eq.~\ref{lISW}. However, from these results it is evident that the late-ISW effect in this model is not as constraining relative to the other observables as it is limited by cosmic variance -- only values of $\sigma_{{\rm T}_\phi}/b_\Lambda^3 \gtrsim 0.30$ are likely to be distinguishable from the standard concordance model prediction.

\section{Discussion and Conclusion} \label{DnC}

In this work we  explored the cosmological signatures of a recently proposed model of quantum cosmology \cite{Alexandre:2022ijm}, where  quantum uncertainties in the physical clock are particularly evident about the time of the transition to accelerated expansion.  One of the features of this model that was identified in the original work is that there is an enhancement in the Hubble parameter around the transition from matter to dark energy domination. Extending  the previous work, we have shown that the growth of structure can strongly constrain the parameter space of this model. In particular, future measurements of the growth factor by DESI will be able constrain the parameter space for values of $\sigma_{{\rm T}_\phi}/b_\Lambda^3 \gtrsim 0.10$. Furthermore, the redshift evolution of $f\sigma_8$ measured from RSD will be a sensitive probe of modification to the cosmology in this theory. While this model can also be constrained by the late-ISW effect, the parameter space that can be probed is limited due to cosmic variance. 

In \cite{Alexandre:2022ijm} it was suggested that the model of quantum cosmology considered here could potentially address the Hubble tension. While we haven't explored this suggestion rigorously through an  MCMC analysis, it seems unlikely given current data that this can be accommodated by the parameter space. Indeed, it is well known that late-time resolutions to the Hubble tension tend to be very well constrained relative to early-Universe modifications - namely those which work on the basis of injecting energy density before recombination to modify the size of the sound horizon \cite{Kamionkowski:2022pkx}. That said, this motivates exploring the possible quantum effects at a transition between matter and radiation domination. Indeed, one of the ``unsettling'' features of models like early dark energy \cite{Karwal:2016vyq} (that are quite successful at addressing the Hubble tension) is their why now problem, i.e. why do they turn on at matter-radiation equality. It would be interesting to study if quantum effects in a model like the one studied here could work as an early Universe solution to the Hubble tension -- though since there will be no reflection in the connection, it is not clear that it will be a nearly as strong of an effect as observed here. At most there will be a refraction/diffraction of the wave. 

While the prospects for the model considered here to address the $H_0$ tension are bleak, it does show some promise for potentially alleviating another growing tension in cosmology, the $S_8$ tension. A particularly interesting feature of this tension is that low-redshift probes ($z \leq 1$) of the clustering of galaxies is lower than that inferred from the CMB and CMB lensing which probes from today to recombination \cite{Chen:2022jzq}. Given that this period is around the same time that dark energy becomes dominate in the Universe, it would be interesting if the observed suppression was somehow connected. Indeed, in the context of the model considered here there is no \textit{why now?} problem, i.e. the suppression in structure growth is a direct consequence of the dark energy-matter transition. Recently, some success was seen with a similar idea \cite{Poulin:2022sgp} where the transition triggers new dynamics in the dark sector to alter the growth of structure and alleviate the tension. We leave further study of the implications of this model for the growth of LSS and the possibility of alleviating the Hubble tension with similar dynamics to those studied here but at matter-radiation equality for future work. 

\section{Acknowledgements}
We thank Stephon Alexander and Marc Kamionkowski for useful conversations. SMK is partially suported by NSF PHY-2014052.

\bibliographystyle{unsrt}
\bibliography{bibo}

\end{document}